\documentclass[journal=jpcltd,manuscript=letter]{achemso}

\usepackage[version=3]{mhchem} 
\usepackage{amsmath}
\usepackage{amssymb}
\usepackage[usenames]{color}
\usepackage{threeparttable}
\usepackage{array}
\usepackage{graphicx}
\usepackage{color}
\usepackage{algorithm}
\usepackage{algorithmic}
\usepackage{multirow}
\usepackage{adjustbox}

\newcommand{\centerfig}[2]{%
\centerline{\includegraphics[#2]{#1}}
}

\author{Marc Delarue}
\email{delarue@pasteur.fr}
\affiliation[Pasteur]
{Unit\'{e} de Dynamique Structurale des Macromol\'{e}cules, 
                  URA 3528 du CNRS, Institut Pasteur, 75015 Paris, France}
          
\author{Patrice Koehl}
\email{koehl@cs.ucdavis.edu}
\affiliation[Davis]
{Department of Computer Sciences and Genome Center,
University of California, Davis, CA 95616, USA}

\author{Henri Orland}
\email{henri.orland@cea.fr}
\affiliation[CEA]
{Institut de Physique Th\'{e}orique, CEA, IPhT, CNRS, URA2306, F-91191 Gif-sur-Yvette, France}

\alsoaffiliation[BCSRC]
{Beijing Computational Science Research Center, No.3 HeQing Road, Haidian District,
 Beijing, 100084, China}
                  
\title[Fast sampling of transition paths]
{Conditioned Langevin Dynamics enables efficient sampling of transition paths}

\begin{document}

\begin{abstract}
We propose a novel stochastic method to generate Brownian
paths conditioned to start at an initial point and end at a given
final point during a fixed time $t_{f}$ under a given potential $U(x)$. These paths are sampled
with a probability given by the overdamped Langevin dynamics. We show
that these paths can be exactly generated by a local {\em Stochastic Partial Differential
Equation (SPDE)}. This equation cannot be solved in general.
We present several approximations that are valid either in the low temperature regime
 or in the presence of barrier crossing. 
We show that this method warrants the generation of statistically independent
transition paths. It is computationally very efficient.
We illustrate the method on the two dimensional Mueller potential as well as
on the Mexican hat potential.
\end{abstract}

 \section{Introduction}

Chemical and biological are controlled by dynamics. Indeed, it is the ability of molecules to change conformations that leads to their activity.
Observing experimentally or predicting functional conformational changes is however a very difficult problem.
At the core of this problem is the fact that a transition between two conformations of a molecule is a rare event compared to the time scale of the internal dynamics of the molecule.
This event is a consequence of random perturbations in the structure of the molecule, drawing its energy from the surrounding heat bath;  it is rare whenever the energy barrier that needs to be crossed is high compared to $k_B T$. 
The transition state theory (TST) offers a framework for studying such rare events \cite{Eyring35, Wigner38}. 
The main idea of the TST is that the transition state is a saddle point of the energy surface for the molecule of interest.
In many cases, the most probable transition path is then simply the minimum energy path (MEP) along that energy surface. 
The TST however is limited to situations in which the potential energy surface is rather smooth; it also assumes that every crossing of the energy barrier through the transition state gives rise to a successful reaction. For systems with a rugged potential energy landscape, or when entropic effects matter, the saddle
points do not necessarily play the role of transition states \cite{Eijnden10}.

To alleviate the shortcomings of the TST, Vanden Eijden and colleagues proposed an alternate view of transitions, the Transition Path Theory (TPT) \cite{VandenEijnden06, Eijnden10, Eijnden14}. At zero temperature the TPT is deemed exact. In principle, it eliminates the need for sampling the transition path ensemble. It also provides a framework for finding the shortest, or most probable transition path between two conformations of a molecule.
As such, it  has served as a touchstone for the development of many path finding algorithms.
Some of those were developed for finding the Minimum Energy Path on the energy surface for a molecule, such as morphing techniques \cite{Kim02, Weiss09}, gradient descent methods \cite{Maragakis05, Zheng07, Tekpinar10, PS10}, the nudged elastic band method (NEB) \cite{Jonsson98, Henkelman00, Sheppard08}  and the string method \cite{E02, Ren05, Ren07,Vanden09, Ren13, Maragliano14}.
Other algorithms are concerned with either finding the Minimum Free Energy Path on the free energy surface for the molecule,\cite{Maragliano06,Pan08,Matsunaga12,Branduardi13}.
or finding the path that minimizes a functional, as implemented in the  Minimum Action Path methods \cite{Olender96, Eastman01, Franklin07, Faccioli06, Vanden08, Zhou08, Carter16}. 
This list is not a comprehensive coverage of all existing techniques for finding transition paths, as this is a very active area of research with many techniques proposed every year. 

Due to the inherent fluctuations underlying the transition phenomenon there are many ways however in which a transition can take place. The methods described above usually generate one path along this transition, the most
``probable" one, where probable refers to minimum energy, free energy, or an action.
Path sampling methods expand upon this view by using this path as a seed to generate a Monte Carlo random walk in the path space of the transition trajectories, 
and thus generate an approximation  of the ensemble of all possible transition paths \cite{Pratt86, Bolhuis02}. All the relevant kinetic and thermodynamic information
related to the transition can then be extracted from the ensemble, such as the reaction mechanism, the transition states, and the rate constants. 
 The main drawback of these methods however  is that they are very time consuming and therefore limited to small systems.
 In addition, they generate highly correlated trajectories in which case the space of sampled trajectories depends strongly on the initial path.
 
In parallel to path sampling methods, much  effort  has  recently  been  dedicated to the development and analysis of Markov State Models
(MSMs) \cite{Chodera07, Bowman09, Pande10}.
MSMs aim at coarse-graining the dynamics of the molecular system via mapping it onto a continuous-time Markov jump process, that  is,  a  process  whose  evolution  involves  jumps  between  discretized  states  representing typical  conformations  of  the original  system.
Much of the recent work focuses on generating those conformations and the dynamics between them, usually using molecular dynamics simulations.
To this day, MSMs remain computationally intensive methods.
 
 In this paper we are concerned with the problem of path sampling. Following preliminary work by one of us \cite{Orland11}, 
 we propose a novel method for generating paths using a Stochastic Partial Differential Equation (SPDE). 
 This equation cannot be solved in general but we were able to find approximations that are valid for different regimes for the dynamics of
 the system. 
 
 The paper is organized as follows. In the next section, we derive the SPDE and describe the different approximations we have implemented to solve this equation.
 In the following section, we show applications to two well-studied 2D problems. Finally, we conclude the paper with a discussion of the extension of the method to study transition pathways for bio-molecular systems.

\section{Theory}

\subsection{Derivation of the bridge equation}

We assume that the system is driven by a force $F(x,t)$
and is subject to stochastic dynamics in the form of an overdamped
Langevin equation.

For the sake of simplicity, we illustrate the method on a one-dimensional
system, the generalization to higher dimensions or larger number of
degrees of freedom being straightforward. We follow closely the presentation
given in Ref. \cite{Orland11}.

The overdamped Langevin equation reads

\begin{equation}
\frac{dx}{dt}=\frac{1}{\gamma}F(x(t),t)+\sqrt{\frac{2 k_{B}T}{\gamma}}\eta(t)\label{eq:langevin}
\end{equation}
where $x(t)$ is the position of the particle at time $t$, driven
by the force $F(x,t)$, $\gamma$ is the friction coefficient, related
to the diffusion coefficient $D$ through the Einstein relation $D=k_{B}T/\gamma$,
where $k_{B}$ is the Boltzmann constant and $T$ the temperature
of the heat bath. In addition, $\eta(t)$ is a Gaussian white noise
with moments given by

\begin{equation}
\langle\eta(t)\rangle=0\label{eq:noise1}
\end{equation}
\begin{equation}
\langle\eta(t)\eta(t')\rangle=\delta(t-t')\label{eq:noise2}
\end{equation}

The probability distribution $P(x,t)$ for the particle to be at point
$x$ at time $t$ satisfies a Fokker-Planck equation \cite{VanKampen92, Zwanzig01},

\begin{equation}
\frac{\partial P}{\partial t}=D\frac{\partial}{\partial x}\left(\frac{\partial P}{\partial x}-\beta FP\right)\label{eq:FP}
\end{equation}
where $\beta=1/k_{B}T$ is the inverse temperature. This equation
is to be supplemented by the initial condition $P(x,0)=\delta(x-x_{0})$,
where the particle is assumed to be at $x_{0}$ at time $t=0$. To
emphasize this initial condition, we will often use the notation $P(x,t)=P(x,t|x_{0},0)$.

We now study the probability over all paths starting at $x_{0}$ at
time $0$ and conditioned to end at a given point $x_{f}$ at time
$t_{f}$, to find the particle at point $x$ at time $t\in[0,t_{f}]$.
This probability can be written
as 
\[
\mathcal{P}(x,t)=\frac{1}{P(x_{f},t_{f}|x_{0},0)}Q(x,t)P(x,t)
\]
where we use the notation 
\[
P(x,t)=P(x,t|x_{0},0)
\]
\[
Q(x,t)=P(x_{f},t_{f}|x,t)
\]

Indeed, the probability for a path starting from $(x_{0},0)$ and
ending at $(x_{f},t_{f})$ to go through $x$ at time $t$ is the
product of the probability $P(x,t|x_{0},0)$ to start at $(x_{0},0)$
and to end at $(x,t)$ by the probability $P(x_{f},t_{f}|x,t)$ to
start at $(x,t)$ and to end at $(x_{f},t_{f})$.

The equation satisfied by $P$ is the Fokker-Planck equation mentioned
above (\ref{eq:FP}), whereas that for $Q$ is the so-called reverse
or adjoint Fokker-Planck equation  \cite{VanKampen92, Zwanzig01} given by

\begin{equation}
\frac{\partial Q}{\partial t}=-D\frac{\partial^{2}Q}{\partial x^{2}}-D\beta F\frac{\partial Q}{\partial x}\label{eq:FTadj}
\end{equation}

It can be easily checked that the conditional probability $\mathcal{P}(x,t)$
satisfies a new Fokker-Planck equation

\[
\frac{\partial\mathcal{P}}{\partial t}=D\frac{\partial}{\partial x}\left(\frac{\partial\mathcal{P}}{\partial x}-\left(\beta F+2\frac{\partial\ln Q}{\partial x}\right)\mathcal{P}\right)
\]

Comparing this equation with the initial Fokker-Planck (\ref{eq:FP})
and Langevin (\ref{eq:langevin}) equations, one sees that it can
be obtained from a Langevin equation with an additional potential
force

\begin{equation}
\frac{dx}{dt}=\frac{1}{\gamma}F+2D\frac{\partial\ln Q}{\partial x}+\sqrt{\frac{2 k_{B}T}{\gamma}}\eta(t)\label{eq:bridge1}
\end{equation}

This equation has been previously obtained using the Doob transform
\cite{Doob57, Fitzsimmons92} in the probability literature and provides a simple recipe
to construct a \emph{generalized bridge}. It generates Brownian paths,
starting at $(x_{0},0)$ conditioned to end at $(x_{f},t_{f})$, with
unbiased statistics. It is the additional term $2D\frac{\partial\ln Q}{\partial x}$
in the Langevin equation that guarantees that the trajectories starting
at $(x_{0},0)$ and ending at $(x_{f},t_{f})$ are statistically unbiased.
This equation can be easily generalized to any number of degrees of freedom.

In the following, we will specialize to the case where the force $F$
is derived from a potential $U(x)$. The bridge equation becomes 
\begin{equation}
\frac{dx}{dt}=-\frac{1}{\gamma}\frac{\partial U}{\partial x}+2D\frac{\partial\ln Q}{\partial x}+\sqrt{\frac{2 k_{B}T}{\gamma}}\eta(t)\label{eq:bridge2}
\end{equation}

In that case, the Fokker-Planck equation corresponding to this modified Langevin equation can be recast into an imaginary
time Schr\"{o}dinger equation \cite{Orland11}, and the probability distribution
function $P$ can be written as 
\begin{equation}
Q(x,t)=P(x_{f},t_{f}|x,t)=e^{-\beta(U(x_{f})-U(x))/2}\langle x_{f}|e^{-H(t_{f}-t)}|x\rangle\label{eq:matrix}
\end{equation}
where $H$ is a "quantum Hamiltonian"
defined by 
\begin{equation}
H=-D\frac{\partial^{2}}{\partial x^{2}}+D\frac{\beta^{2}}{4}V(x)
\end{equation}
and the potential $V$ by 
\begin{equation}
V=\left(\frac{\partial U}{\partial x}\right)^{2}-2k_{B}T\frac{\partial^{2}U}{\partial x^{2}}
\end{equation}

We denote by $M$ the matrix element of the Euclidian Schr\"{o}dinger
evolution operator 
\begin{equation}
M(x,t)=\langle x_{f}|e^{-H(t_{f}-t)}|x\rangle\label{eq:mat}
\end{equation}

Using eq.(\ref{eq:matrix}) for $Q$, one can write equation (\ref{eq:bridge2})
as

\begin{equation}
\frac{dx}{dt}=2\frac{k_{B}T}{\gamma}\frac{\partial}{\partial x}\ln M(x,t)+\sqrt{\frac{2 k_{B}T}{\gamma}}\eta(t)\label{eq:bridge3}
\end{equation}

We see on the above form that when $t\to t_{f}$, the matrix element
$M(x,t)$ converges to $\delta(x_{f}-x)$, and it is this singular
attractive potential which drives all the paths to $x_{f}$ at time
$t_{f}$.

\subsection{Transition paths}

The bridge equations (\ref{eq:bridge2}) or (\ref{eq:bridge3}) can be solved exactly in a certain number of cases \cite{Majumdar15}. However in general, for systems with many degrees of freedom, the functions $Q(x,t)$ or $M(x,t)$ cannot be computed exactly and one has to resort to some approximations.
In the following, we will be mostly interested in problems of energy
or entropy barrier crossing, which are of utmost importance in many
chemical, biochemical, or biological reactions. 

The matrix element $M(x,t)$ can be written as a Feynman path integral

\begin{equation}
M(x,t)=\int_{x(t)=x}^{x(t_{f})=x_{f}}\mathcal{D}x\:\exp\left(-\int_{t}^{t_{f}}d\tau\left(\frac{1}{4D}\left(\frac{dx}{d\tau}\right)^{2}+\frac{D\beta^{2}}{4}V(x(\tau))\right)\right)\label{mat1}
\end{equation}

The free case is defined as

\begin{eqnarray}
M_{0}(x,t)&=&P_{0}(x_{f},t_{f}|x,t) \nonumber \\
&=& \int_{x(t)=x}^{x(t_{f})=x_{f}}\mathcal{D}x\:\exp\left(-\int_{t}^{t_{f}}d\tau\left(\frac{1}{4D}\left(\frac{dx}{d\tau}\right)^{2}\right)\right)\nonumber \\
&=&\left(\frac{1}{4\pi D(t_{f}-t)}\right)^{1/2}e^{-\frac{\left(x_{f}-x\right)^{2}}{4D(t_{f}-t)}} \label{mat0} 
\end{eqnarray}
where $P_{0}$ is the probability distribution for a free particle.

Equation (\ref{mat1}) can be rewritten as

\[
M(x,t)=M_{0}(x,t)\langle \exp\left(-\frac{D\beta^{2}}{4}\int_{t}^{t_{f}}d\tau V(x(\tau))\right)\rangle_{0}
\]
where the expression $\langle...\rangle_{0}$ denotes the expectation value with
the Brownian measure $P_{0}$.

The convexity of the exponential function implies the Jensen
inequality \cite{Jensen1906}, which states that for any operator $A$ and any probability
measure, one has

\begin{equation}
\langle e^{-A}\rangle \geq e^{-\langle A \rangle}\label{jensen}
\end{equation}

Equality occurs when the probability is a $\delta-$function;
it is thus a good approximation when the operator $A$ has small fluctuations.Taking
$A$ to be 
\begin{equation}
A=\frac{D\beta^{2}}{4}\int_{t}^{t_{f}}d\tau V(x(\tau)) \label{A}
\end{equation}
we have

\begin{equation}
M(x,t)\simeq M_{0}(x,t)\exp\left(-\frac{D\beta^{2}}{4}\int_{t}^{t_{f}}d\tau \langle V(x(\tau)) \rangle_{0}\right)\label{jensen1}
\end{equation}

Using the expression
\[
\langle V(x(\tau)) \rangle_{0}=\frac{1}{M_{0}(x,t)}\int dz P_{0}(x_{f},t_{f}|z,\tau) V(z) P_{0}(z,\tau|x,t),
\]
after some calculations, we obtain

\[
\langle V(x(\tau)) \rangle_{0}=\left(\frac{\theta_1+\theta_2}{4\pi D\theta_{1}\theta_{2}}\right)^{1/2}\int dz\exp\left(-\frac{\theta_1 + \theta_2}{4 D \theta_{1}\theta_{2}}\left(z-\frac{x_{f}\theta_{2}+x\theta_{1}}{\theta_1+\theta_2}\right)^{2}\right)V(z)
\]
where

\[
\theta_{1}=t_{f}-\tau,\;\theta_{2}=\tau-t
\]
After a change of variable, this expression becomes

\[
\langle V(x(\tau)) \rangle _{0}=\int\frac{dz}{\left(2\pi\right)^{1/2}}\exp\left(-\frac{z^{2}}{2}\right)V\left(X+\sqrt{\frac{2D\theta_{1}\theta_{2}}{\theta_1+\theta_2}}z\right)
\]
where

\[
X=\frac{x_{f}\theta_{2}+x\theta_{1}}{\theta_1+\theta_2}
\]
and the constrained Langevin equation (\ref{eq:bridge3}) becomes

\begin{eqnarray}
\frac{dx}{dt}&=&\frac{x_{f}-x}{t_{f}-t}\nonumber \\
&-&\frac{(D\beta)^{2}}{2}\int_{t}^{t_{f}}d\tau\left(\frac{t_{f}-\tau}{t_{f}-t}\right)\int\frac{dz}{\left(2\pi\right)^{1/2}}\exp\left(-\frac{z^{2}}{2}\right)\frac{\partial}{\partial X}V\left(X+\sqrt{\frac{2D(t_{f}-\tau)(\tau-t)}{(t_{f}-t)}}z\right) \nonumber \\
&+&\sqrt{\frac{2 k_{B}T}{\gamma}}\eta(t)\label{approx1}
\end{eqnarray}
or, after the change of variable $u=\frac{\tau-t}{t_f-t}=\frac{\theta_2}{\theta_1 + \theta_2}$,
\begin{eqnarray}
\frac{dx}{dt}&=&\frac{x_{f}-x}{t_{f}-t} \nonumber \\
&-&\frac{(D\beta)^{2}}{2}(t_f-t)\int_{0}^{1}du (1-u) \int\frac{dz}{\left(2\pi\right)^{1/2}}\exp\left(-\frac{z^{2}}{2}\right)\frac{\partial}{\partial X}V\left(X+\sqrt{2D(t_{f}-t)u(1-u)}z\right) \nonumber \\
&+&\sqrt{\frac{2 k_{B}T}{\gamma}}\eta(t)\label{approx1a}
\end{eqnarray}
where
\begin{equation}
X= x_f u + x (1-u)
\end{equation}

Integration by part with respect to $z$ yields the equivalent form


\begin{eqnarray}
\frac{dx}{dt}&=&\frac{x_{f}-x}{t_{f}-t} \nonumber \\
&-&\left(\frac{D}{2}\right)^{3/2} \beta^2 \sqrt{t_f-t} \int_{0}^{1}du \sqrt \frac{1-u}{u}\int\frac{dz}{\left(2\pi\right)^{1/2}}\exp\left(-\frac{z^{2}}{2}\right)zV\left(X+\sqrt{2D(t_{f}-t)u(1-u)}z\right) \nonumber \\
&+&\sqrt{\frac{2 k_{B}T}{\gamma}}\eta(t)\label{approx2}
\end{eqnarray}
which does not require the cumbersome evaluation of $\partial V / \partial X$.
These forms require an integration over the Gaussian variable $z$
which can be performed by sampling this variable
\[
\int\frac{dz}{\left(2\pi\right)^{1/2}}\exp\left(-\frac{z^{2}}{2}\right)F(z)\simeq\frac{1}{M}\sum_{i}F(z_{i})
\]
where the $M$ variables $z_{i}$ are Gaussian variables (with zero
average and unit variance).

However, as we have seen, the approximation (\ref{jensen1}) is valid
if the exponent $A$ does not fluctuate too much over the trajectories
relevant to the transition. There are two cases when this approximation
can be further simplified and where the $z$-integral can be avoided:

\begin{enumerate}

\item Low temperature

In that case, since $D=k_{B}T/\gamma$, diffusion
is small, thus $V$ can be approximated as $(\frac{\partial U}{\partial x})^{2}$. In addition, the term $\sqrt{2D(t_{f}-t)u(1-u)} z$ in (\ref{approx1a}) is small compared to $X$ and can be neglected.
Equation (\ref{approx1}) can be simplified to
\begin{eqnarray}
\frac{dx}{dt}=\frac{x_{f}-x}{t_{f}-t}-\frac{(D\beta)^{2}}{2}(t_f-t)\int_{0}^{1}du (1-u) \frac{\partial}{\partial X}
\left(\frac{\partial}{\partial X}U\left(X\right)\right)^{2}
+\sqrt{\frac{2 k_{B}T}{\gamma}}\eta(t)\label{approx3}
\end{eqnarray}

\item Barrier crossing

According to Kramers theory, the total transition time $\tau_{K}$
(waiting + crossing) scales like the exponential of the barrier height
$\Delta E^{*}$ while it has been shown that the crossing time (Transition
Path Time) $\tau_{c}$ scales like the logarithm of the barrier $\Delta E^{*}$ \cite{Gopich06, Kim15, Carlon16}.
We have thus $\tau_{c}<<\tau_{K}$.

As discussed before, the barrier crossing time is
very short compared to the Kramers time. Therefore the transition
trajectories are very weakly diffusive, and are thus almost
ballistic. Consequently, we have $\sqrt{2Dt_{f}}\ll |x_{f}-x_{0}|$
and again we can neglect the $z$ term in $V$. Equation (\ref{approx1})
becomes
\begin{equation}
\label{approx4}
\frac{dx}{dt}=\frac{x_{f}-x}{t_{f}-t}-\frac{(D\beta)^{2}}{2}(t_f-t)\int_{0}^{1}du (1-u) \frac{\partial  V(X)}{\partial X}
+\sqrt{\frac{2 k_{B}T}{\gamma}}\eta(t)
\end{equation}

\end{enumerate}

All the equations described above are easily generalized to any number of particles in any dimension, interacting with any many-body potentials.
They are integro-differential stochastic Markov equations,  as the variable $X$ depends only on the stochastic variable $x(t)$. One can generate many independent trajectories by integrating these equations with different noise histories $\eta(t)$.
To test the validity of the main approximation (\ref{jensen1}), one should compute the variance $(\Delta A)^2 = \langle A^2\rangle-\langle A \rangle^2$ of the random variable $A$ in eq.(\ref{A}) over all the trajectories generated. Computing the correction to the Jensen inequality, it is easily seen that the approximation is reliable provided 
\begin{equation}
R= \frac{(\Delta A)^2}{2 |\langle A \rangle| }<<1 \label{criterion}
\end{equation}

\subsection{Simulation Time}

For barrier crossing, what simulation time $t_f$ should be used? 
Obviously, for any initial and final state $x_i$ and $x_f$, there is a set of Langevin trajectories which make the transition in any given time $t_f$.
If the time $t_f$ is very short compared to the typical time scales of large motions of the system, there is a small number of such trajectories, since they require a very specific noise history. As a result, the approximations presented above are reliable and the factor $R$ is much smaller than 1. However, this is not a very interesting regime, as trajectories are driven by the boundary conditions.
If we are interested in simulating transition paths, the time $t_f$ should obviously be larger than the typical TPT $\tau_c$. Indeed, if $t_f$ is smaller than $\tau_c$, we will simulate paths driven by the final state. On the other hand, if $t_f$ is too large, then we will also simulate part of the waiting time in the wells, where fluctuations are large (except maybe at  low temperature). Therefore, in order to simulate transition paths as accurately as possible, one should use a simulation time $t_f$ larger than the typical TPT $\tau_c$, but not much larger.

\section{Results}

We now illustrate these concepts on two examples: the Mueller potential,
and the Mexican hat potential.

\subsection{The Mueller potential}

\begin{figure}
	\centerfig{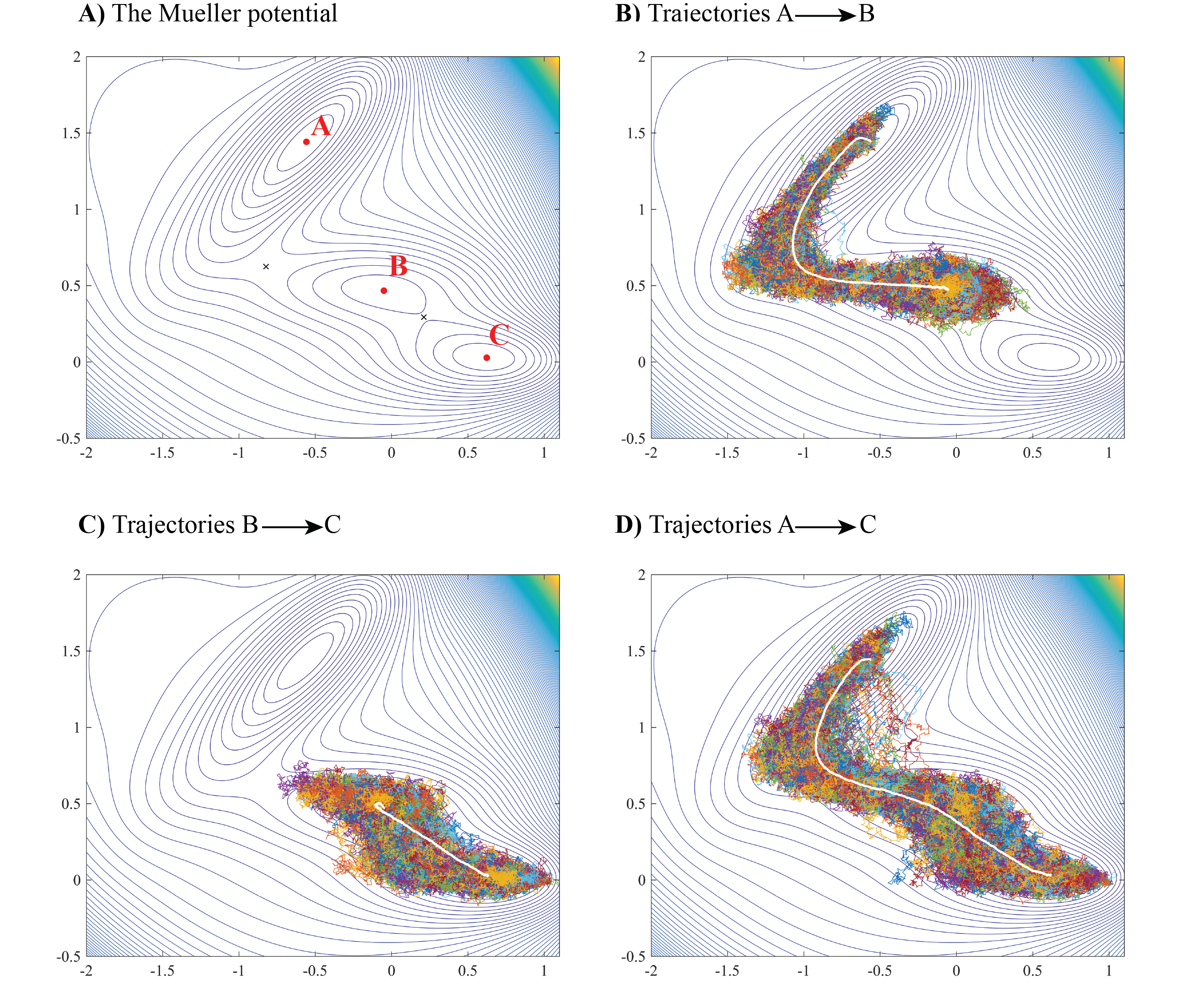}{height=6in}
	\caption{\textbf{Langevin bridge trajectories on the Mueller potential}. (A) Contour plot of the Mueller potential, with the three minima labeled A, B, and C, and the two saddle points between those minima indicated with an x. 500 converged trajectories between the minima A and B (B), B and C (C), and A and C (D). The unweighted mean trajectories are shown in white.}.
	\label{fig:mueller}
\end{figure}

The Mueller potential is a standard benchmark potential to check the validity of methods for generating transition paths.
It is a two dimensional potential given by
\begin{equation}
U(x,y)= \sum_{i=1}^4 A_i \exp \left( a_i(x-x_i^0)^2+b_i(x-x_i^0)(y-y_i^0)+c_i(y-y_i^0)^2 \right) \label{Mueller}
\end{equation}
with
\begin{equation}
A=(-200,-100,-170,15) \ \ \ a=(-1,-1,-6.5,0.7)\ \ \ b=(0,0,11,0.6)
\end{equation}
This potential has 3 local minima denoted by A,B,C, separated by two barriers (Fig.\ref{fig:mueller}). The effective potential $V(x,y)$ can be calculated analytically, as well as its gradient. Equations (\ref{approx1}), (\ref{approx3}) and (\ref{approx4}) can easily be solved numerically. We display only the trajectories generated by (\ref{approx4}). The simulation time $t_f$ is chosen so that we observe a small waiting time around the initial as well as the final point, namely $t_f=0.15$. 
We use 50 points for the integration over $u$.
We display a sample of 500 trajectories obtained from eq. (\ref{approx4}) with $t_f=0.15,\ dt=10^{-4},\ D=1$ at temperature $T=5$. We can compute the average trajectory as well as its variance. These trajectories are displayed on Fig. \ref{fig:mueller}, where we plot the AB, BC and AC trajectories.

\begin{figure}
	\centerfig{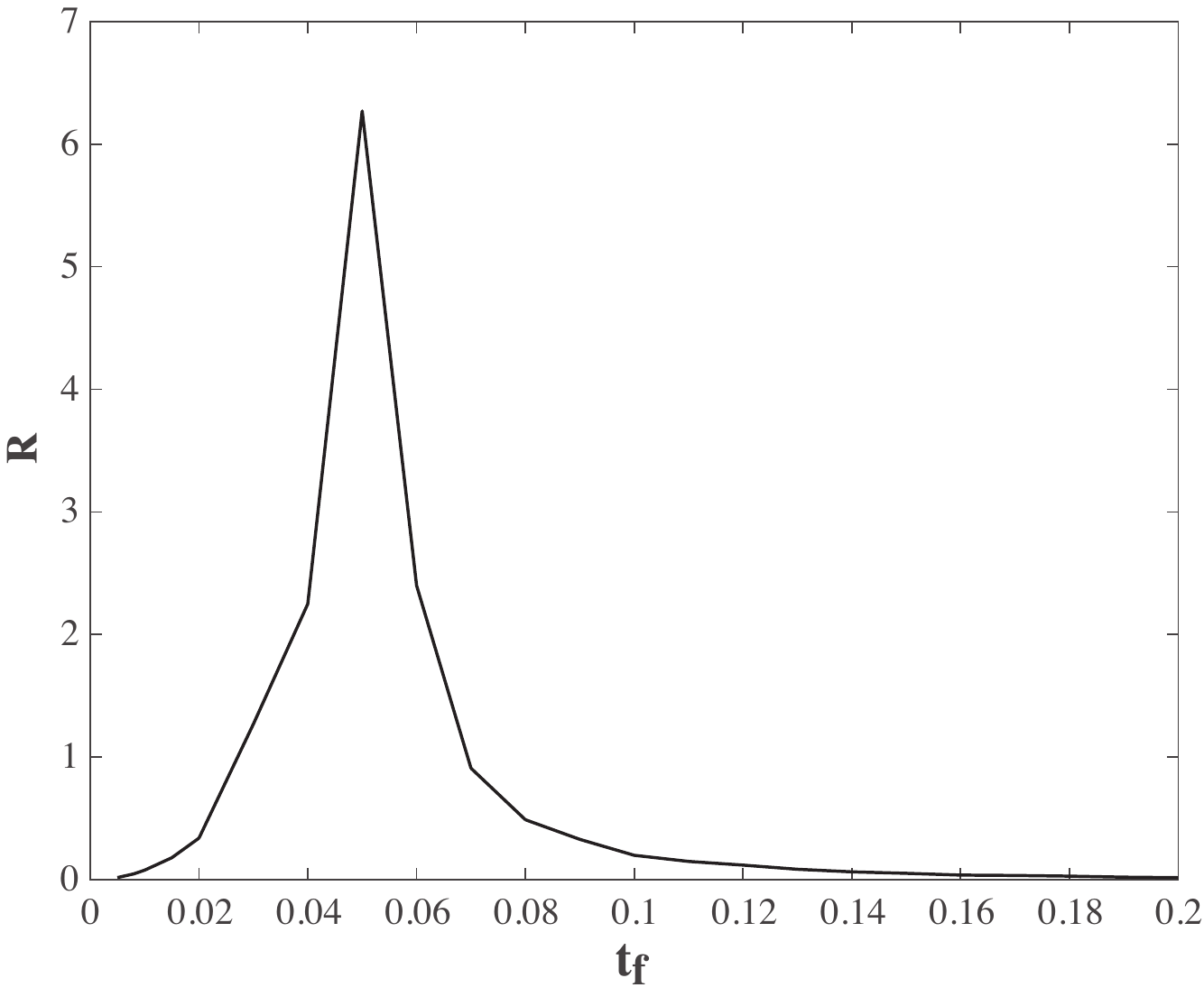}{height=2in}
	\caption{The quality factor $R$ plotted as a function of the total duration of the transition $t_f$, for the AB trajectories.}
	\label{fig:rplot}
\end{figure}

To assess the quality of the approximations, we check the criterion (\ref{criterion}). For the trajectories AB, we obtain $R \approx 5.3 10^{-2}$, $R \approx 0.68$ for BC and $R \approx 1.13$ for AC. Therefore, the approximation is quite reliable for the AB trajectories, but less for the others. In fact, it is instructive to study the accuracy of the method when varying $t_f$. For that matter, in Fig. \ref{fig:rplot}, we plot the factor $R$ as a function of $t_f$, for the AB transition. We see that for both small and large $t_f$, the factor $R$ is small, with a maximum at $t_f \approx 0.05$. For small $t_f$, the trajectories fluctuate around the straight line trajectory joining A to B through high barriers (see Fig. \ref{fig:mueller}A). For large $t_f$, the trajectories fluctuate around the potential energy valley joining A to B. As $t_f$ increases from small values, the ensemble of trajectories include trajectories going through the high barrier and through the valley, and at $t_f\approx 0.05$, there is a strong mixing of both types of trajectories, giving rise to a large value of $R$. When $t_f$ increases further, the trajectories going through the barrier disappear from the ensemble, and only valley trajectories remain, yielding a decrease of $R$.

\subsection{The Mexican hat potential}

\begin{figure}
	\centerfig{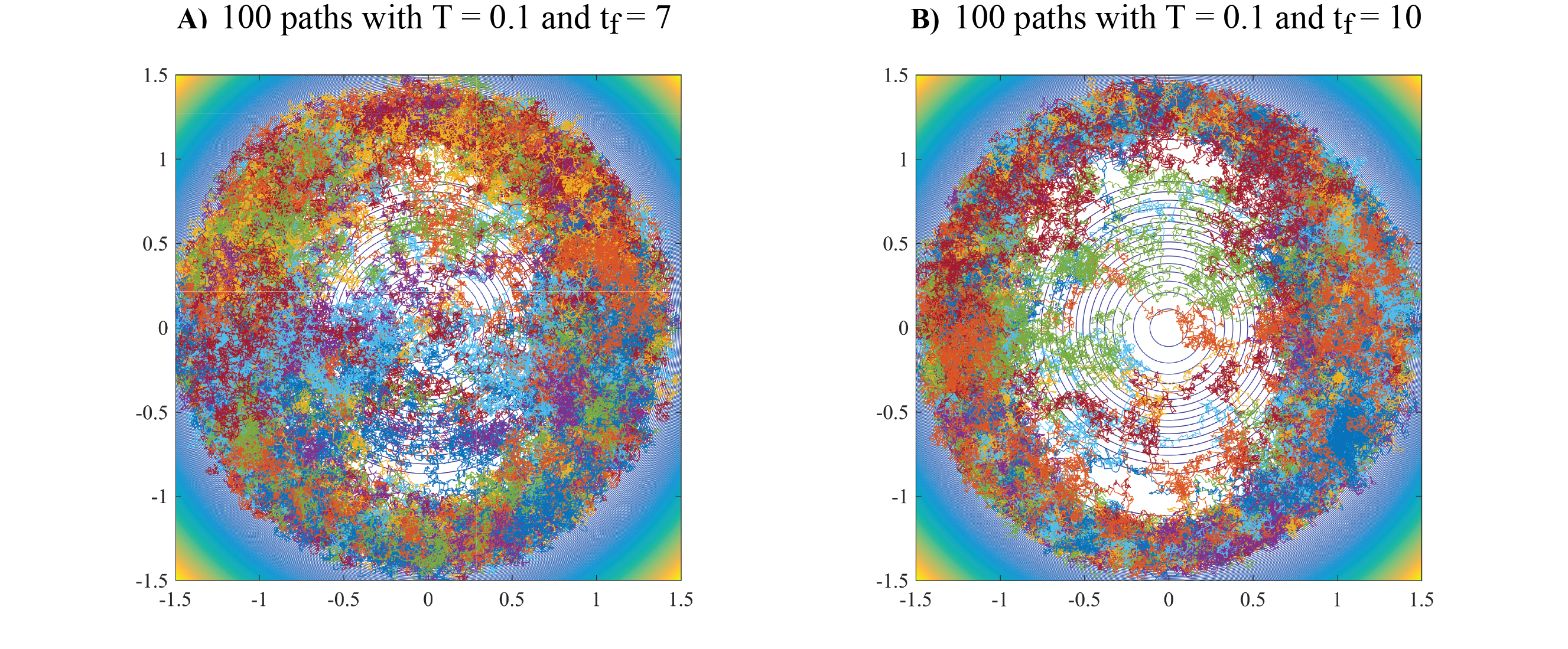}{height=3in}
	\caption{\textbf{Langevin bridge trajectories on the Mexican hat potential}. (A) 100 converged trajectories, all starting at $(-1,0)$ and ending at $(1,0)$, and generated at temperature $T=0.1$ with a duration $t_f=7$ . Note that with this short transition time, many trajectories go through the barrier region.  (B) Same as in (A), but with $t_f$ now set to 10. Most of the trajectories now follow the circle of minima; those trajectories are nearly equally divided into two groups, those that follow the upper side of the circle (44), and those that follow the lower side (47) (see text for details).}.
	\label{fig:sombrero}
\end{figure}

The potential of the Mexican hat is given by
\begin{equation}
U(x,y)=\frac{1}{4}(x^2+y^2-1)^2
\end{equation}
and has therefore a circle of minima for $x^2+y^2=1$ with $U=0$ and a maximum at $(0,0)$ with energy $U=1/4$.
Again we solve equations (\ref{approx4}), using 50 points for the $u$ integral. Given the small barrier of this potential $\Delta U=1/4$, we go to low temperature. On Fig.\ref{fig:sombrero}A, we plot 100 trajectories, generated at temperature $T=0.1$, starting at $(-1,0)$ and all ending at $(1,0)$. The total time is $t_f=7$ and the time step is $dt=10^{-4}$. The quality criterion (\ref{criterion}) gives $R=0.345$. The trajectories divide into three dominant groups, those that take a northern route (30), those that take a southern route (40) along the circle of minima, and those that go directly through the energy barrier (30). The distribution into those three groups was decided based on  the mean value $Y_{mean}$ for the $y$ coordinates along the trajectories. If we take a longer duration, the fraction of trajectories that go through the central barrier decreases. For example, for  $t_f=10$, there are only 9 of those trajectories , as seen on Fig. \ref{fig:sombrero}B. The quality criterion is then $R=0.266$.

\section{Conclusions}

In summary, we show here a novel method to generate ab initio
transition path trajectories using a formalism called Conditioned Langevin dynamics.
The most crucial parameter of the theory is the total length of the simulation,
which must be carefully chosen so as to allow short waiting times around both the initial
and final states. We define a quality criterion R that can be calculated \textit{a posteriori}
to see if the approximation made to solve the underlying PDE is justified or not.
Here we have tested the method on simple one- or two-dimensional potentials, but
our aim is to use it in more complicated situations, such as conformational transitions in proteins.\\ 

Proteins are small biopolymers (up to a few hundred amino-acids)
that do not stay in one of their two states (e.g. native or denatured, open or closed, apo or holo in the allosteric picture), 
but rather make unfrequent stochastic transitions between them. 
They are usually represented in a coarse-grained manner using one or two beads per amino-acid.
The picture which emerges is that of the system staying for a long time in one of the
minima and then making stochastically rapid transitions to the
other minimum. It follows that for most of the time, the system
performs harmonic oscillations in one of the wells, which can be described
by normal mode analysis. Rarely, there is a very short but interesting
physical phenomenon, where the system makes a fast transition between minima. 
This picture has been confirmed by single
molecule experiments, where the waiting time in one state can be measured,
although the time for crossing is so short
that it cannot be resolved \cite{Chung12}. This scenario has also been confirmed
recently by very long millisecond molecular dynamics simulations which
for the first time show spontaneous thermal folding-unfolding events \cite{Lindorff11}.

It is possible to model this behavior using a an energy function that is based on a simplified Elastic Network Model
that is a mixture of two Elastic Networks centered around each of the two states of the macromolecule \cite{Maragakis05, Das14}.
We will present the results of the Conditioned Langevin Equation applied to this situation in a forthcoming paper.

\section{Acknowledgements}

MD acknowledges financial support from the Agence Nationale de la Recherche (ANR) through the program Bip-Bip. 

\bibliography{langevin3}

\end{document}